\documentclass[a4paper,twoside]{article}
\usepackage{amssymb}
\usepackage{amsmath}
\usepackage{amsfonts}

\begin{document}

\title{\Large\bf{Disappearance of the Black Hole Singularity\\ in Quantum Gravity}} 

 \author{\\ Leonardo Modesto \\[1mm] \em\small{Centre de Physique
 Th\'eorique de Luminy, 
 Universit\'e de la M\'editerran\'ee,
 }\\[-1mm] \em\small{ 
 Case 907, F-13288
 Marseille, EU.}\\[-1mm] \em \small{Dipartimento di Fisica
 dell'Universit\`a di Torino, INFN - Sez.  di Torino,
 }\\[-1mm] \em\small{ 
 via P. Giuria 1, I-10125 Torino, EU} \\[-1mm] }
 \date{\ } \maketitle

\begin{abstract}

We apply techniques recently introduced in quantum cosmology to the
Schwarzschild metric inside the horizon and near the black hole
singularity at $r=0$.  In particular, we use the quantization
introduced by Husain and Winkler, which is suggested by Loop Quantum
Gravity and is based on an alternative to the Schr\"odinger
representation introduced by Halvorson.  Using this quantization
procedure, we show that the black hole singularity disappears and
spacetime can be dynamically extended beyond the classical
singularity.

\end{abstract}
 
\section*{Introduction}

A remarkable result of loop quantum cosmology \cite{Boj} is the
disappearance of the initial cosmological singularity present in the
classical theory.  The main results of loop quantum gravity
\cite{book}, indeed, are the quantization of area and volume partial
observables \cite{Partial}, which suggest that in the complete theory
there cannot be spacetime points with infinity matter density.  If
this is correct, the quantum theory should control all classical
singularities of general relativity.  In this work, we apply
techniques analogous to the ones used in loop quantum cosmology to
study the $r=0$ singularity in the interior of a Schwarzschild black
hole.

In particular, we use the non-Schr$\ddot{\mbox{o}}$dinger procedure of
quantization introduced by Halvorson \cite{Fonte.Math} and utilized in
quantum cosmology by Husain and Winkler \cite{Fonte}.  We focus on the
Schwarzschild solution inside the horizon and near the singularity. 
We use the method introduced in \cite{Thie} to express $1/r$, and
therefore the curvature invariant $\mathcal{R}_{\mu \nu \rho \sigma}
\, \mathcal{R}^{\mu \nu \rho \sigma} = 48 M^2 G_N^2/r^6$, in terms of
the volume operator.  Following \cite{Thie}, we write the Hamiltonian
constraint as well in terms of the volume.  This allows us to express
the quantum evolution equation as a difference equation for the
coefficients for the physical states, and to completely control the
singularity.

The paper is organized as follow.  In the first section we briefly
recall the properties of the Schwarzschild solution for $r<2M G_N$,
namely inside the horizon.  As well known, here the temporal and
spatial (radial) coordinate exchange their role.  In the second
section we study the classical dynamics of a very simple model giving
this solution.  The Hamiltonian constraint depends on a single
variable, and its classical solution yields the Schwarzschild metric
inside the horizon, in the new temporal variable.  In the third
section we quantize the system using the
non-Schr$\ddot{\mbox{o}}$dinger procedure of quantization of
references \cite{Fonte.Math,Fonte}.  In particular, we show that the
singularity in $r = 0$ is resolved in quantum gravity and that the
Hamiltonian constraint acts like a difference operator, as in loop
quantum cosmology.

\section{The Schwarzschild Solution Inside the Horizon}

Consider the Schwarzschild solution 
\begin{eqnarray}
ds^2 = - \Big(1 - \frac{2 M G_N}{r} \Big) dt^2 + \frac{dr^2}{\Big(1 -
\frac{2 M G_N}{r} \Big)} + r^2 (\sin^2 \theta d\phi^2 + d \theta^2)
\end{eqnarray}
for $r<2MG_{N}$.  This metric describe spacetime inside the horizon of
a Schwarzschild black hole.  The coordinate $r$ is timelike and the
coordinate $t$ is spatial; for convenience we rename them as $r \equiv
T$ and $t \equiv r$ with $T \in ] 0 , 2 M G_N[$ and $r \in ] - \infty
, + \infty[$. The metric reads then
\begin{eqnarray}
ds^2 = - \frac{dT^2}{\Big(\frac{2 M G_N}{T} -1 \Big)} + \Bigg(\frac{2
M G_N}{T} -1 \Bigg) dr^2 + T^2 (\sin^2 \theta d\phi^2 + d \theta^2).
\end{eqnarray}
We eliminate the coefficient of $dT^2$ by defining a new temporal
variable $\tau$ via
\begin{equation}
d \tau = \frac{dT}{\sqrt{\frac{2 M G_N}{T} -1}} .
\label{tempo}
\end{equation}
The integration gives 
\begin{equation}
\tau = - \sqrt{T(2 M G_N -T)} + 2 M G_N \arctan \Bigg(\sqrt{\frac{T}{2
M G_N -T}}\Bigg) + const.
\label{solcamb}
\end{equation}
We take $const = 0$ because $\mbox{lim}_{T \rightarrow 0} \, \tau(T) =
const$.  The function $T = T(\tau)$ is monotonic and convex, thus $\tau
\in ] 0 , 2 M G_N \pi /2[$.  In this new temporal variable the metric
becomes
\begin{eqnarray}
ds^2 = - d \tau^2+ \Bigg(\frac{2 M G_N}{T(\tau)} -1 \Bigg) dr^2 +
T(\tau)^2 (\sin^2 \theta d\phi^2 + d \theta^2).
\end{eqnarray}
We introduce two function $a^2(\tau) \equiv \frac{2 M G_N}{T(\tau)} -
1$ and $b^2(\tau) \equiv T^2(\tau)$ and redefine $\tau \equiv t$.  The
metric reads
\begin{eqnarray}
ds^2 = - d t^2+ \Bigg(\frac{2 M G_N}{b(t)} -1 \Bigg) dr^2 +
b(t)^2 (\sin^2 \theta d\phi^2 + d \theta^2).
\label{metricb}
\end{eqnarray}
Notice that a metric written in terms of two functions $a(t)$ and
$b(t)$ with the form
\begin{eqnarray}
ds^2 = - dt^2 + a^2 (t) dr^2 + b^2 (t)  (\sin^2 \theta d\phi^2 + d \theta^2)
\label{metricab}
\end{eqnarray}
is the metric of an homogeneous, anisotropic space with spatial
section of topology $\mathbf{R} \times \mathbf{S}^2$.  In our case,
$a(t)$ is a function of $b(t)$, $a=a(b(t))$.

\section{Classical Theory}

The complete action for gravity can be written in the form
\begin{eqnarray}
S = \frac{M_p^2}{16 \pi} \int d^3 x dt N h^{1/2} \Big[ K_{ij} K^{ij} -K^2 + ^{(3)}\mbox{R}\Big] 
\label{Action}
\end{eqnarray}
If we specialize this for metrics of the form (\ref{metricab}), the
action becomes \cite{cava}
\begin{eqnarray}
S & = & -  \frac{M_p^2}{16 \pi}\int dt  \int_0^R dr \int_0^{2 \pi} d \phi \int_0^{2 \pi} d \theta \sin \theta
 \, b^2 \, a \Bigg[2 \, \frac{\dot{b}^2}{b^2} + 4 \, \frac{\dot{a} \, \dot{b}}{a \, b} - \frac{2}{b^2} \Bigg] = \nonumber \\
    & = & - \frac{M_p^2 R}{2}\int dt \Big[a \, \dot{b}^2 + 2 \, \dot{a} \, \dot{b} \, b - a\Big].
\label{Action.Mini} 
\end{eqnarray}
Recalling from (\ref{metricb}) that the two functions $a(t)$ and $b(t)$
are not independent, and satisfy
\begin{eqnarray}
a^2(t) = \frac{2 M G_N}{b(t)} - 1,
\end{eqnarray}
we can write the action in terms of a single function
\begin{eqnarray}
S = \frac{M_p^2 R}{2}\int dt \Bigg[\frac{\sqrt{b}}{\sqrt{2 M G_N}}
\Big(1 - \frac{b}{2 M G_N}\Big)^{-\frac{1}{2}} \, \dot{b}^2 + 2
\,\frac{\sqrt{b}}{\sqrt{2 M G_N}} \Big(1 - \frac{b}{2 M G_N}\Big)^{\frac{1}{2}}
\Bigg].
\label{Action.Mini.ridotta} 
\end{eqnarray}
Now we calculate the Hamiltonian, which is also the Hamiltonian
constraint (see Appendix B).
The momentum is
\begin{eqnarray}
p = M_p^2 R \frac{\sqrt{b}}{\sqrt{2 M G_N}} \Big(1 - \frac{b}{2 M
G_N}\Big)^{-\frac{1}{2}} \, \dot{b},
\end{eqnarray}  
and so 
\begin{eqnarray}
H = p \, \dot{b} - L = \Bigg( \frac{p^2}{2 M_P^2 R} - \frac{M_P^2
R}{2} \Bigg) \Bigg[ \frac{\sqrt{2 M G_N}}{\sqrt{b}} \Big(1 -
\frac{b}{2 M G_N}\Big)^{\frac{1}{2}} \Bigg].
\label{Hamiltonian}
\end{eqnarray}
We can now show that the Hamiltonian constraint produce the correct
classical dynamics.  We express the Hamiltonian constraint in terms of
$\dot{b}$
\begin{eqnarray}
H = M_P^2 R \left[ \frac{\dot{b}^2}{\sqrt{\frac{2MG_N}{b} -1}} -
\sqrt{\frac{2 MG_N}{b} -1 } \, \right] = 0
\end{eqnarray}
The solution is 
\begin{eqnarray}
\dot{b}^2 = \Big( \frac{2 MG_N}{b} -1  \, \Big) 
\end{eqnarray}
and this is exactly the equation (\ref{tempo}) with solution (\ref{solcamb}) that reproduces the 
Schwarzschild metric.
 
We now introduce an approximation.  In the quantum theory, we will be
interest in the region of the scale the Planck length $l_{p}$ around
the singularity.  We assume that the Schwarzschild radius $r_{s}
\equiv 2 M G_N$ is much larger than this scale, and that $b(t) =
T(t)$. In this approximation we can write
\begin{eqnarray}
1 -  \frac{b}{2 M G_N} \sim 1
\end{eqnarray}
and $H$ becomes 
\begin{eqnarray}
H = \Bigg( \frac{p^2}{2 M_P^2 R} - \frac{M_P^2}{2} \Bigg)
\frac{\sqrt{2 M G_N}}{\sqrt{b}}.
\label{Hamiltonian.approx}
\end{eqnarray}
The volume is 
\begin{eqnarray}
V = \int dr \, d \phi \, d \theta \, N \, h^{1/2} = 4 \pi R a b^2 = 4
\pi R \sqrt{2 M G_N} \, b^{3/2} \, \sqrt{1 - \frac{b}{2 M G_N}};
\label{Volume}
\end{eqnarray}
in the previously approximation 
\begin{eqnarray}
V =  4 \pi R \sqrt{2 M G_N} \, b^{3/2} \equiv l_o \, b^{3/2}.
\label{Volume.approx}
\end{eqnarray}
The canonical pair is given by $b \equiv x$ and $p$, with Poisson
brackets $\{x,p\} = 1$.  

We now assume that $x \in \mathbb{R}$ (and introduce the absolute
value where appropriate).  This choice it not correct classically,
because for $ b \equiv x =0$ we have the singularity.  But it allows
us to open the possibility that the situation be different in the
quantum theory.  We introduce an algebra of classical observables and
we write the quantities of physical interest in terms of those
variables.  We are motivated by loop quantum gravity to use the
fundamental variables $x$ and
\begin{eqnarray}
U_{\gamma}(p) \equiv \mbox{exp} \Big(\frac{8 \pi G_N \gamma}{L} \, i \,  p\Big) 
\end{eqnarray}   
where $\gamma$ is a real parameter and $L$ fixes the unit of length. 
The parameter $\gamma$ is necessary to separate the momentum point in
the phase space.  (Choosing $\gamma/L = 1$ we obtain the some value of
$U$ for $p$ and $p + 2 \pi n$).  This variable can be seen as the
analog of the holonomy variable of loop quantum gravity.

A straightforward calculation gives
\begin{eqnarray}
&& \{x , U_{\gamma}(p)\} = 8 \pi G_N \frac{i \, \gamma}{L} U_{\gamma}(p) \nonumber \\
&& U_{\gamma}^{-1} \{ V^n , U_{\gamma} \} = l_0^n \,
U_{\gamma}^{-1} \{ |x|^{\frac{3n}{2}} , U_{\gamma} \} = i \,8 \pi
G_N \, l_0^n \, \frac{\gamma}{L} \frac{3 n}{2}Ê\mbox{sgn}(x)
|x|^{\frac{3n}{2} -1}
\label{Poisson.Volume}
\end{eqnarray}
These formulas allow us to express inverse powers of $x$ in terms of a
Poisson bracket, following Thiemann's trick \cite{Thie}.  As we will see below,
the volume operator has zero as an eigenvalue, therefore so we must
take $n \geqslant 0$ for the second equation to be well define din the
quantum theory.  On the other hand, if we want that the power of $x$
on the right hand side be negative we need $n \leqslant 2/3$.  The
choice $n = 1/3$ gives
\begin{eqnarray}
\frac{\mbox{sgn}(x)}{\sqrt{|x|}} = - \frac{2 L  \,
i}{(8 \pi G_N) l_0^{\frac{1}{3}} \gamma} \, U_{\gamma}^{-1} \{ V^{\frac{1}{3} },
U_{\gamma} \}. 
\label{unosux}
\end{eqnarray}
We use this relation in the next section to write physical operators. 
We are interested to the quantity $\frac{1}{|x|}$ because classically
this quantity diverge for $|x| \rightarrow 0$ and produce the
singularity.  We are also interested to the Hamiltonian constraint and
the dynamics and we will use (\ref{unosux}) for writing the
Hamiltonian.

\section{Quantum Theory}

We construct the quantum theory proceeding in analogy with the
procedure used in loop quantum gravity.  The first step is the choice
of an algebra of classical functions to be represented as quantum
configuration operators.  We choose the algebra generated by the
functions
\begin{eqnarray}
W(\lambda) = e^{i \lambda x/L}
\end{eqnarray}
where $\lambda \in \mathbb{R}$. The algebra consists of all function of the form 
\begin{eqnarray}
f(x) = \sum_{j=1}^{n} c_j e^{i \lambda_j x/L}
\end{eqnarray}
where $c_j \in \mathbb{C}$, and their limits with respect to the sup
norm.  This is the algebra $AP(\mathbb{R})$ of the \emph{almost
periodic functions} over $\mathbb{R}$ .  The algebra $AP(\mathbb{R})$
is isomorphic to $C(\bar{\mathbb{R}}_{Bohr})$, the algebra of
continuous functions on the Bohr-compactification of $\mathbb{R}$. 
This suggests to take the Hilbert space $L_2(\bar{\mathbb{R}}_{Bohr},
d \mu_0)$, where $d \mu_0$ is the Haar measure on
$\bar{\mathbb{R}}_{Bohr}$.  With this choice the basis states in the
Hilbert space are
\begin{eqnarray}
&& |\lambda \rangle \equiv |e^{i \lambda x /L} \rangle \nonumber \\
&& \langle \mu | \lambda \rangle = \delta_{\mu, \lambda}
\end{eqnarray}
The action of the configuration operators $\hat{W}(\lambda)$ on the
basis is defined by
\begin{eqnarray}
\hat{W}(\lambda) | \mu \rangle = e^{i \lambda \hat{x} /L} | \mu \rangle = e^{i \lambda \mu } |\mu \rangle
\end{eqnarray}
These operators are weakly continuous in $\lambda$.  This implies the
existence of a self-adjoint operator $\hat{x}$, acting on the basis
states according to
\begin{equation}
\hat{x} |\mu \rangle = L \mu |\mu \rangle
\label{xoperator}
\end{equation}
Next, we introduce the operator corresponding to the classical
momentum function $U_{\gamma} = e^{i \gamma p/L}$.  We define the
action of $\hat{U}_{\gamma}$ on the basis states using the definition
(\ref{xoperator}) and using a quantum analog of the Poisson bracket
between $x$ and $U_{\gamma}$
\begin{eqnarray}
&& \hat{U}_{\gamma} |\mu \rangle = | \mu - \gamma \rangle \nonumber \\
&& \left[ \hat{x} , \hat{U}_{\gamma} \right] = - \gamma L \hat{U}_{\gamma}
\end{eqnarray} 
Using the standard quantization procedure $[ \, , \, ] \rightarrow i
\hbar \{ \, , \, \}$, and using (\ref{Poisson.Volume}) we obtain
\begin{eqnarray}
&& - \gamma = \hbar (8 \pi G_N) \frac{i \gamma}{L} \nonumber \\
&& L = \sqrt{8 \pi} \, l_p
\end{eqnarray}

\subsection{Volume operator and disappearance of the singularity}

Near the singularity we can use the approximation
(\ref{Volume.approx}).  The action of the volume operator on the basis
states is
\begin{eqnarray}
\hat{V} | \mu \rangle = l_0 |x|^{\frac{3}{2}} | \mu \rangle = l_0 |L
\mu|^{\frac{3}{2}} |\mu \rangle. 
\end{eqnarray}

Recall that the dynamics is all in the function $b(t)$, which is equal
to the the radial Schwarzschild coordinate inside the horizon $b(t) =
T(t)$.  The function $b(t)$ generated by the dynamics is monotonic and
convex.  The important point is that $b(t = 0) = 0$ and this is the
Schwarzschild singularity.  We now show that the term $\frac{2 M
G_N}{b(t)}$ does not diverge in the quantum theory and therefore there 
is no singularity in the quantum theory.

We use the relation (\ref{unosux}) and we promote the Poisson brackets
to commutators.  In this way we obtain (for $\gamma = 1$) the operator
\begin{eqnarray}
 \widehat{\frac{1}{|x|}} = \frac{1}{2 \pi l_p^2 l_0^{\frac{1}{3}}} \left(
 \hat{U}^{-1} \left[ \hat{V}^{\frac{1}{3}} , \hat{U} \right] \right)^2. 
\end{eqnarray}
The action of this operator on the basis states is
  \begin{eqnarray}
 \widehat{\frac{1}{|x|}} \, | \mu \rangle = \sqrt{\frac{2}{ \pi l_p^2}}
 \left( | \mu |^{\frac{1}{2}} - |\mu -1|^{\frac{1}{2}}\right)^2 \, |
 \mu \rangle. 
\end{eqnarray}
We can now see that the spectrum is bounded from below and so we have
not singularity in the quantum theory.  In fact, for example, the
curvature invariant
\begin{eqnarray}
\mathcal{R}_{\mu \nu \rho \sigma} \, \mathcal{R}^{\mu \nu \rho \sigma}
= \frac{48 M^2 G_N^2}{r^6} \equiv \frac{48 M^2 G_N^2}{T^6} = \frac{48
M^2 G_N^2}{T(t)^6} \equiv \frac{48 M^2 G_N^2}{b(t)^6}
\end{eqnarray}
is finite in quantum mechanics in fact the eigenvalue of $1/|x|$ for
the state $|0 \rangle$ corresponds to the classical singularity and in
the quantum case it is $\sqrt{2/\pi l_p^2}$, which is the largest
possible eigenvalue.  For this particular value the curvature
invariant it is not infinity
 \begin{eqnarray}
\widehat{\mathcal{R}_{\mu \nu \rho \sigma} \, \mathcal{R}^{\mu \nu
\rho \sigma}} |0 \rangle = \widehat{\frac{48 M^2 G_N^2}{|x|^6}} |0
\rangle = \frac{384 M^2 G_N^2}{\pi^3 l_P^6} |0 \rangle. 
\end{eqnarray}
On the other hand, for $|\mu | \rightarrow \infty$ the eigenvalues go
to zero, which is the expected behavior of $1/|x|$ for large $|x|$.

\subsection{Hamiltonian Constraint}

We now study the quantization of the Hamiltonian constraint near the
singularity, in the approximation (\ref{Hamiltonian.approx}).  There
is no operator $p$ in quantum representation that we have chosen,
hence we choose the following alternative representation for $p^2$. 
Consider the classical expression
\begin{eqnarray}
p^2 = \frac{L^2}{(8 \pi G_N)^2} \mbox{lim}_{\gamma \rightarrow 0}
\left( \frac{2 - U_{\gamma} - U_{\gamma}^{-1}}{\gamma^2} \right).
\end{eqnarray}
We have can give a physical interpretation to $\gamma$ as $\gamma =
l_p / L_{phys}$, where $L_{Phys}$ is the characteristic size of the
system. Using this, we write the Hamiltonian constraint as
\begin{eqnarray}
\hat{H} = \frac{A_1}{l_0^{1/3}} \left[ \hat{U}_{\gamma} +
\hat{U}_{\gamma}^{-1} - (2 - A_2) \, 1\, \right] \mbox{sgn}(x) \left(
\hat{U}^{-1} \left[ \hat{V}^{\frac{1}{3}} , \hat{U} \right] \right)
\end{eqnarray}
where $A_1 = \frac{L^3}{{(8 \pi G_N})^{5/2} \gamma ^3 M_P^2 \, R \, l_0^{1/3} \hbar }$
and $A_2 = \frac{R^2 \gamma^2}{8 \pi l_P^2}$.  The action of $\hat{H}$
on the basis states is
\begin{eqnarray}
\hat{H} | \mu \rangle = \mathcal{C} \, \mathcal{V}_{\frac{1}{2}}(\mu)
\left[ | \mu - \gamma \rangle + | \mu + \gamma \rangle - (2 -
\mathcal{C}^{\prime}) | \mu \rangle \right],
\end{eqnarray}
where $\mathcal{C} = A_1 L^{1/2}$ and $\mathcal{C}^{\prime} \equiv
A_2$, and
\begin{equation}
\mathcal{V}_{\frac{1}{2}}(\mu) = \left\{ \begin{array}{cc} - \big|
|\mu - \gamma|^{1/2} - |\mu|^{1/2} \big| \hspace{0.5cm} \mbox{for}
\hspace{0.5cm} \mu \neq 0 \\
 |\gamma|^{1/2} \hspace{1cm} \mbox{for} \hspace{0.5cm} \mu = 0 \\
 \end{array}\right.
\label{sistem3}
\end{equation}
If we calculate the action of $\hat{H}$ and $1/|x|$ on the state of
zero volume eigenvalue we obtain
\begin{eqnarray}
&& \hat{H} |0 \rangle = \mathcal{C} |\gamma|^{\frac{1}{2}} \left[ | - \gamma \rangle + |\gamma \rangle -
(2 - \mathcal{C}^{\prime}) |0 \rangle \right] \nonumber \\
&& \widehat{\frac{1}{|x|}} \, | 0 \rangle = \sqrt{\frac{2}{ \pi l_P^2}} \,
|0 \rangle.
\end{eqnarray} 
This finite value of ${\frac{1}{|x|}}$ can be interpreted as the
effect of the quantization on the classical singularity.  

We now study the solution of the the Hamiltonian constraint.  The
solutions are in the $\mathcal{C}^{\star}$ space that is the dual of
the dense subspace $\mathcal{C}$ of the kinematical space
$\mathcal{H}$.  A generic element of this space is
\begin{eqnarray}
\langle \psi | = \sum_{\mu} \psi(\mu) \langle \mu |.
\end{eqnarray}
The constraint equation $\hat{H} |\psi \rangle = 0$ is now interpreted
as an equation in the dual space $\langle \psi | \hat{H}^{\dag}$; from
this equation we can derive a relation for the coefficients
$\psi(\mu)$
\begin{eqnarray}
\mathcal{V}_{\frac{1}{2}} (\mu + \gamma) \, \psi(\mu + \gamma) +
\mathcal{V}_{\frac{1}{2}} (\mu - \gamma) \, \psi(\mu - \gamma) - (2 -
\mathcal{C}^{\prime})\, \mathcal{V}_{\frac{1}{2}}(\mu) \, \psi(\mu) =
0.
\label{difference}
\end{eqnarray}
This relation determines the coefficients for the physical dual state. 
We can interpret this states as describing the $quantum$ $spacetime$
near the singularity.  From the difference equation (\ref{difference})
we obtain physical states as combinations of a countable number of
components of the form $\psi(\mu + n \gamma) |\mu + n \gamma \rangle $
($\gamma \sim l_P/L_{Phys} \sim 1$); any component corresponds to a
particular value of volume, so we can interpret $\psi(\mu + \gamma)$
as the wave function describing the black hole near the singularity at
the time $\mu + \gamma$.  A solution of the Hamiltonian constraint
corresponds to a linear combination of black hole states for
particular values of the volume or equivalently particular values of
the time.

\section*{Conclusions}

We have applied the quantization procedure of \cite{Fonte} to the case
of the Schwarzschild singularity.  This procedure is alternative to
the Schr$\ddot{\mbox{o}}$dinger quantization and it is suggested by
loop quantum cosmology.  The main results are:
\begin{enumerate}
\item The classical black hole singularity near $r \sim 0$, which in
our coordinate is $b(t) \equiv T(t) \sim 0 $, disappears from the
quantum theory.  Classical divergent quantities are bounded in the
quantum theory.  For instance:
\begin{eqnarray}
\mathcal{R}_{\mu \nu \rho \sigma} \, \mathcal{R}^{\mu \nu \rho \sigma}
= \frac{48 M^2 G_N^2}{b(t)^6} \hspace{0.5cm} \rightarrow
\hspace{0.5cm} \widehat{\mathcal{R}_{\mu \nu \rho \sigma} \,
\mathcal{R}^{\mu \nu \rho \sigma}} |0 \rangle = \widehat{\frac{48 M^2
G_N^2}{|x|^6}} |0 \rangle = \frac{384 M^2 G_N^2}{\pi^3 l_P^6} |0
\rangle. \nonumber
\end{eqnarray}
\item The quantum hamiltonian constraint gives a discrete difference
equation for the coefficients of the physical states.
\end{enumerate}

It is interesting to observe that beyond the classical singularity the
function $b \equiv x$ is negative.  One can speculate, extrapolating
the form of the metric that ``on the other side" of the singularity
there is no horizon: a black hole and a naked singularity are connected \cite{easson}.

\section*{Acknowledgements}

I am grateful to Carlo Rovelli, and Eugenio Bianchi for many important
and clarifying discussions.  This work was partially supported by the
Fondazione Angelo Della Riccia


\section*{Appendix A}
We give here the explicit form of some tensors used in the paper.  The
spatial diagonal metric tensor is
\begin{equation}
h_{ij} = \left(\begin{array}{ccc}
                    a^2(t) &     0                                   &    0  \\
                         0    &     b^2(t) \sin^2 \theta    &    0  \\
                         0    &     0                                   &    b^2(t)
\end{array}   \right).
\end{equation}
The inverse spatial metric tensor is
\begin{equation}
h^{ij} = \left(\begin{array}{ccc}
                    a^{-2}(t) &     0                                   &    0  \\
                         0    &     b^{-2}(t) \sin^{-2} \theta    &    0  \\
                         0    &     0                                   &    b^{-2}(t)
\end{array}   \right).
\end{equation}
The extrinsic curvature is $ K_{ij} = -\frac{1}{2} \frac{\partial h_{ij}}{\partial t}$, and so
\begin{equation}
K_{ij} = \left(\begin{array}{ccc}
                   - a \, \dot{a} &     0                                             &    0  \\
                         0            &     -b \, \dot{b} \sin^2 \theta       &     0  \\
                         0            &     0                                              &   -b \, \dot{b}
\end{array}   \right).
\end{equation}
\begin{eqnarray}
&& K \equiv K_{ij} h^{ij} = - \Bigg( \frac{\dot{a}}{a} + 2 \, \frac{\dot{b}}{b} \Bigg) \nonumber \\
&& K_{ij} K^{ij} =  \frac{\dot{a}^2}{a^2} + 2 \, \frac{\dot{b}^2}{b^2} \nonumber \\
&& K_{ij} K^{ij}  - K^2 = - \Bigg(\frac{2 \, \dot{b}^2}{b^2} + 4 \, \frac{\dot{a} \, \dot{b}}{a b} \Bigg)
\end{eqnarray}
The Ricci curvature for the space section is 
\begin{eqnarray}
^{(3)} \mbox{R} = \frac{2}{b^2}
\end{eqnarray}

\section*{Appendix B} 
In this appendix we report the Hamiltonian for our system and we show that reproduces 
the correct equation of motion. We can start from the Hamiltonian
\begin{eqnarray}
H = \Bigg( \frac{p^2}{2 M_P^2 R} - \frac{M_P^2
R}{2} \Bigg) \Bigg[ \frac{\sqrt{2 M G_N}}{\sqrt{b}} \Big(1 -
\frac{b}{2 M G_N}\Big)^{\frac{1}{2}} \Bigg],
\label{Hamiltonian2}
\end{eqnarray}
and calculate the Hamilton equation for $b$ ($\dot{b} = \frac{\partial H}{\partial p}$) 
\begin{eqnarray}
\dot{b} = \frac{p}{M_P^2 R} \sqrt{\frac{2 MG_N}{b} - 1} \, .
\label{bdot}
\end{eqnarray}
At this point using the constraint $H = 0$ with (\ref{bdot}), we obtain 
\begin{eqnarray}
\dot{b}^2 = \Big( \frac{2 MG_N}{b} -1  \, \Big),
\end{eqnarray}
that is the equation of motion for $b(t)$ that reproduce the Schwarzschild solution.

\end{document}